\documentclass[manuscript]{acmart}
\AtBeginDocument{%
  }

\usepackage{booktabs}
\usepackage{longtable}
\usepackage{enumitem}
\usepackage{makecell}
\setcopyright{acmlicensed}
\usepackage{xcolor}
\copyrightyear{2018}
\begin{document}

\title{Does AI Save Time on Product Design? A Randomized Controlled Experiment of AI Prompt-to-Design Workflows}

\author{Remy Stewart}
\email{rstewart@figma.com}
\affiliation{%
  \institution{Figma}
  \city{San Francisco}
  \state{California}
  \country{USA}
}

\author{Olabode Anise}
\email{oanise@figma.com}
\affiliation{%
  \institution{Figma}
  \city{San Francisco}
  \state{California}
  \country{USA}
}

\author{Andrew Hogan}
\email{ahogan@figma.com}
\affiliation{%
  \institution{Figma}
  \city{San Francisco}
  \state{California}
  \country{USA}
}

\author{Augustus Griffin}
\email{agriffin@figma.com}
\affiliation{%
  \institution{Figma}
  \city{San Francisco}
  \state{California}
  \country{USA}
}

\renewcommand{\shortauthors}{Stewart et al.}

\begin{abstract}
AI tools for digital product design now offer prompt-to-design capabilities, allowing designers and their non-designer colleagues to create prototypes through conversational workflows with large language models (LLMs). While these tools promise time savings, experimental evidence in product design remains limited compared with evidence from software engineering. We conducted a randomized controlled trial with 50 product designers and 50 product managers to evaluate prospective time savings from leveraging Figma Make in design work. Participants attempted three standardized design tasks with or without access to Figma Make. Among participants who completed the study tasks, access to Figma Make was associated with approximately 20\% shorter completion times, with larger gains among product managers. Our findings suggest that prompt-to-design tools may enable product managers to further contribute to design work, while the benefits for professional designers may be task dependent.

\end{abstract}

\begin{CCSXML}
<ccs2012>
<concept>
<concept_id>10003120.10003121.10011748</concept_id>
<concept_desc>Human-centered computing~Empirical studies in HCI</concept_desc>
<concept_significance>500</concept_significance>
</concept>
<concept>
<concept_id>10003120.10003121.10003122.10003334</concept_id>
<concept_desc>Human-centered computing~User studies</concept_desc>
<concept_significance>500</concept_significance>
</concept>
<concept>
<concept_id>10003120.10003123.10011760</concept_id>
<concept_desc>Human-centered computing~Systems and tools for interaction design</concept_desc>
<concept_significance>500</concept_significance>
</concept>
</ccs2012>
\end{CCSXML}

\ccsdesc[500]{Human-centered computing~Empirical studies in HCI}
\ccsdesc[500]{Human-centered computing~User studies}
\ccsdesc[500]{Human-centered computing~Systems and tools for interaction design}

\keywords{Design Software, AI Time Savings, AI Productivity, AI-Assisted Design, Randomized Controlled Trial}

\received{20 February 2007}
\received[revised]{12 March 2009}
\received[accepted]{5 June 2009}

\maketitle

\section{Introduction}
A core value proposition behind why individuals should adopt AI-powered tools is to automate routine tasks and save time. The promise of increased productivity has helped drive both the development of AI tools and their adoption in the workplace. In a 2026 survey, 79\% of corporate organizations reported adopting generative AI to some degree \cite{sajadieh2026}. Prior research identifies employee productivity as a partial mediator between AI capabilities and organizational performance and links workforce productivity to organizational readiness and innovative practices \cite{kassa2025, watts2025}. 

Digital product design is one of the many professions rapidly adopting AI tools, with 91\% of designers reporting at least weekly usage in another 2026 survey \cite{designerfund2026}. HCI research examining why designers integrate AI into their workflows consistently identifies perceived productivity gains, defined by metrics such as tasks completed, workflows automated, and overall time saved \cite{luo2025, naqvi2025, cha2026}. Concurrently, other HCI research has identified how AI tools can jeopardize productivity, as well as how productivity alone does not engage with AI's impacts on other dimensions such as design values, innovation, and collaboration  \cite{takaffoli2024, li2024, uusitalo2024, khan2025, zulu2026}. Productivity gains can often come with tradeoffs on overall design quality via dimensions such as originality and legibility \cite{wadinambiarachchi2024}. Challenges with AI prompting and the stochasticity of outputs can also make tasks harder to complete \cite{subramonyam2025}. 

Although designers often cite time savings as a reason for adopting AI tools, there has been limited work to quantify the time savings from AI adoption on everyday design tasks. Experimental research pertaining to time-saving measurement has instead focused on evaluating AI coding tools such as GitHub Copilot, Claude Code, and Cursor on software engineers' task completion times \cite{peng2023, paradis2025, mohamed2026, chen2026}. Randomized controlled trials (RCTs) have been a primary method within this body of research due to their particular strength towards mitigating confounders. This study asks a question previously examined in software engineering by employing the same RCT design: Does using AI save design professionals time in their everyday work?

AI tools for design are evolving rapidly alongside advances in AI research and engineering.  With these advances, two notable trends have emerged: the rise of conversational  or "vibe-based" prompt-to-design tools and lower barriers to participation in design work for non-designers. Prompt-to-design tools are one of the latest developments in integrating AI into design software. Earlier integrations focused on narrower tasks, such as auto-completion or image editing, or relied on external chatbots like ChatGPT. By contrast, tools such as Lovable, Figma Make, and v0 can generate fully deployable designs and their supporting code through iterative conversations with users, extending AI’s role across design and software development. These prompt-to-design tools may reduce the expertise required to contribute to  design work by shifting fluency with concepts such as design systems and interaction prototyping from the individual to the supporting AI tool. This has enabled other job roles to participate further in design projects alongside product designers, such as with product managers (PMs), UX researchers, and software engineers \cite{li2026, kobiella2026}. 

This study examines two potential benefits of prompt-to-design tools: reducing the time required for everyday design tasks and enabling non-designers to contribute to design work. We investigate these hypothesized benefits through an experiment testing Figma Make, a prompt-to-design tool integrated into the widely used Figma design platform \cite{designerfund2026}. Our research questions are as follows: 

\begin{enumerate}[label=\textbf{RQ\arabic*:}]
    \item Does using AI tooling such as Figma Make on common design work save time in minutes compared to not using AI tools? 
    \item Are there differences on the impact of time spent on design work with or without AI tools for designers compared to non-designers?
    \item How does using AI tools in design work additionally affect design quality, ease of work, and platform usability?
\end{enumerate}

We engaged with these three research questions by conducting a moderated between-subjects randomized controlled trial in which participants attempted a standardized series of edits on a social media UI design. Our results demonstrated a 20\% improvement in cumulative time savings among participants who completed the study tasks, as well as a 16\% gain in rated design task ease and a 15\% improvement in perceived Figma usability. The most sizable gains were driven by product managers over product designers, with product designers not reporting a statistically significant aggregate time savings gain on their own. Our findings indicate that product manager time savings occurred on less complex design tasks, while time savings for product designers only emerged on the most complex tested task. We discuss these findings in relation to ongoing advances in AI design tools.

\section{Related Work}

\subsection{Measuring AI Tooling's Impact on Productivity}
Experimental research of AI's impact on productivity has been led by measuring time savings for programming by software engineers, as well as additional industries such as customer service and consulting \cite{noy2023, weber2024, paradis2025, dellacqua2026, demirer2026}. The magnitude and direction of AI's impact have been mixed. While studies such as Peng et al. found reductions of up to 55\% in time to task completion with 2023's AI tools, other work such as Becker et al. found productivity regressions by 19\% when considering AI in 2025 \cite{peng2023, becker2025}. Measured effect sizes have often varied by level of experience in a particular profession. Bryjolfsson et al. found that the aggregate 15\% increase in issues resolved per hour was predominantly driven by gains among newer and lower-skilled employees, while more experienced workers saw minimal gains. Similarly, Cui et al. reported a  25\% increase in completed programming tasks, with larger gains among less experienced developers \cite{brynjolfsson2025, cui2025}.

Additional dimensions such as the study setting, employed AI tool, and experiment design have been found to substantially impact variability in measured productivity outcomes within randomized controlled trials \cite{maier2026}. The heterogeneity of these measurements reflects the nuances around when and how AI can enable productivity gains. A subset of studies on AI's impact on productivity within coding further complicates the narrative of whether AI provides a net positive productivity impact. He et al.'s work on programming supported by Cursor's AI agent found short-term task completion time improvements, but a reduction in overall code quality that slows down long-term velocity due to accumulated tech debt \cite{he2026}. Afroz et al. flagged similar tradeoffs between faster task completion enabled by AI that is then offset by increased demands to verify and test AI output \cite{afroz2026}.

Similar dynamics play out in the world of AI tool adoption for design and productivity, even if scholarship has not predominantly focused on experimental approaches. Randomized experiments testing AI's impact on design work have concentrated on alternative measures such as creativity \cite{kumar2025}, divergent thinking \cite{xu2025}, ability to problem reframe \cite{shin2025}, or when users decide to use or not use AI given trade offs on latency and error rates \cite{qiao2025}. Some scholars contest whether net productivity gains actually occur when considering how AI introduces new complexities and management requirements \cite{simkute2025, parsons2026}, while others note that Gen AI's impact on productivity and efficiency is most holistically understood through multiple alternative dimensions including craftsmanship and user empowerment \cite{sun2025, swift2026}. 

HCI scholarship building from interviews, surveys, and other methodologies provides diverse insights regarding both how designers seek out AI tools for productivity gains and the challenges they encounter when using AI that reduces work efficiency. AI tools are highlighted in early design project stages for enabling fast brainstorming and iterations, assisting designers to overcome the "blank canvas problem" and mock up many initial ideas quickly \cite{tholander2023, palani2024, zhou2026, yi2026, cavallin2026}. Designers testify to the advantage of delegating routine, time intensive design work to AI tools, allowing them to create valuable designs more efficiently \cite{takaffoli2024, luo2025}. However, incorrect or poor quality output, insufficient human-centered context to create optimal designs, and cognitive overload caused by managing AI tools can reduce AI's productivity benefits \cite{hong2023, tankelevitch2024, zhu2025, cha2025}. The above heterogeneous effects of AI tooling on design productivity demonstrate how multiple concurrent dynamics influence whether AI saves designers time. 

\subsection{The Rise of Prompt-to-Design AI Tools}
The innovations of AI within design have shifted from Gen AI creative content tooling such as Midjourney and Stable Diffusion, to LLM-based chatbots via ChatGPT, to the latest iterations being AI agents capable of advanced cross-platform work and vibe coding interfaces through conversational workflows. Vibe coding is a particular LLM-user interaction paradigm within software engineering where users delegate code creation and edits predominantly to the AI tool, leaning instead into project innovation and creativity via AI-human co-collaboration \cite{pimenova2025, sarkar2025, fawzy2026}. 

Our work investigates time savings from prompt-to-design tools branching from vibe coding that designers and their collaborators are now rapidly adopting. Hwang and Kang present the concept of "vibe design" to refer to many of these AI products as a designated framework that incorporates both design and user testing agentic support \cite{hwang2026}, but our conceptualization in this research is broader. These AI tools rely on prompt-based conversations between humans and LLMs in chat panel interfaces where chains of prompts lead to iterative and exploratory design creations. These tools create both UI and UX components of digital designs, as well as supporting code to enable deployment of created designs into production environments. The product design software market is now flushed with AI applications from this ecosystem such as Lovable, Replit, v0, Google Stitch, Figma Make, Framer, Bolt, and Claude Design. 

Research on prompt-to-design AI tools so far indicates the benefits these products provide for quickly exploring and iterating on designs. Li et al.'s study participants noted the collapsing of design stages into single conversational workflows through adopting these tools as a particular advantage that reduced cognitive burden and freed up bandwidth for creativity \cite{li2026}. Kobiella et al.'s participants testify to these tools' efficiency with quickly advancing initial ideas to working prototypes, additionally handling code implementation that designers may not have technical backgrounds on \cite{kobiella2026}.

An additional trend within the rise of prompt-to-design products is how they contribute to blurring the lines of which job roles are participating in design work. Many of these products intentionally aim for non-designers as key expansion audiences. AI design tools speed up creating code for new design system artifacts for engineers to refine and integrate into established code bases \cite{irawati2025}. Engineers, data scientists, and product managers at Microsoft have designed their own productivity assistants as early adopters of recent Gen AI tooling enabled at the company \cite{naik2025}. Non professional designers report the ability to move from passive observers to active collaborators with both designers and AI through adopting end-to-end AI tooling \cite{ma2026}. We are therefore interested in measuring whether AI enabled time savings occur in design work beyond just product designers to reflect the expanding scope of who contributes to design empowered by AI design products. 

This study focuses on measuring time savings from Figma Make as a popular AI design tool. 68\% of professional product designers surveyed in 2026 use Figma AI including Figma Make \cite{designerfund2026}. Figma Make created interactive prototypes and the code underlying generated designs with the notable advantage of producing outputs that source from designer's preexisting design systems - the diverse components of design interfaces consistently referenced throughout an organization - which design organizations tend to already house within Figma libraries. Other Figma Gen AI enabled features such as Figma widgets have been the tools of focus in prior HCI work \cite{feng2025}, as well as understanding how designers integrate external AI design tools with their preexisting Figma based workflows \cite{chen2025, irawati2025}. This study builds from these previous investigations by studying Figma Make as one of the latest AI prompt-to-design features launched directly for the Figma platform. Figma Make is additionally targeted to beyond designers as key audiences, with product managers as a main secondary user base \cite{webster2026}.  In sum, we aim to quantify productivity impacts for both product designers and product managers using Figma Make as defined by time to task completion through a laboratory setting randomized controlled experiment.

\section{Methodology}
\subsection{Experiment Design}

We conducted a between-subjects randomized experiment to evaluate whether incorporating Figma Make into a design workflow reduced the time required to complete UI modifications when compared to manual editing in Figma Design. First, participants completed an eligibility questionnaire that additionally collected demographic information. Eligible participants were then randomly assigned to the treatment or control condition. Participants in both conditions attempted the same three design tasks using identical starting designs and reference images of the intended outputs. After the design tasks, participants completed an exit survey assessing perceived Figma usability, self-reported design quality, and prior AI tool usage. 

\subsubsection{Recruitment}
We collaborated with the external research firm MeasuringU to recruit participants and moderate the study. MeasuringU recruited 50 product designers and 50 product managers from their preexisting research panels. Within each role, we randomly assigned participants to the treatment or control condition, resulting in 25 participants from each role in each condition. The sample size was identified by referencing prior AI time savings experiments' samples \cite{peng2023, paradis2025}, as well as forecasting minimum detectable effect size scenarios and modeling required samples for adequate statistical power.  Eligibility was determined from prospective participants' self-reported responses to the recruitment screener. Participants qualified for the study if they had at least two years of experience as a product designer or product manger, viewed or edited a Figma file at least once per month, currently worked in a role involving at least one digital product, and had never been employed by Figma. All participants signed consent forms and received a monetary honorarium. 

\subsubsection{Experiment Flow}
To test the study's central research questions, we designed an experiment focused on modifying and extending an existing UI design. Participants in both the treatment and control conditions attempted the same three tasks, using the same starting designs and reference images of the intended outputs. We opted to use a social media application as the shared scenario for each of the tasks because we expected participants, irrespective of industry background, to be familiar with its basic features. We sought guidance from multiple professional product designers regarding which tasks would arise when iterating on a social media design prototype. We used this input to select categories, which became the distinct tasks of the experiment scenario:

\begin{enumerate}
    \item \textbf{Visual styling}: UI changes to the base design, such as adjusting color palettes, text, or images. Task 1 asked participants to convert the starting light-mode version of the design into a dark-mode variant. 
    \item \textbf{Interface modification}: Adding content to an existing interface element. Task 2 asked participants to add a "Help and Support" option to a dropdown settings menu.
    \item \textbf{Interaction prototyping}: Creating interactions and animations within the existing design. Task 3 asked participants to create a comment flyout that appeared when the social media post's 'Comments' icon was clicked.
\end{enumerate}

All participants attempted the three tasks in the same fixed order: dark-mode conversion, menu modification, and comment flyout creation. Figure 1 illustrates the workflow for each task. Each task's starting and final design is provided in Appendix A. For each task, participants received an initial design in an editable Figma frame whose components participants could directly edit. We provided the target design as a static image rather than an editable Figma frame so that the internal components of the finished design  were not readily available to participants. Participants edited their designs in a designated workspace between the starting frame and the reference image. Edits from earlier tasks did not carry over to later tasks.

\begin{figure}[t!]
  \centering
  \includegraphics[width=\linewidth]{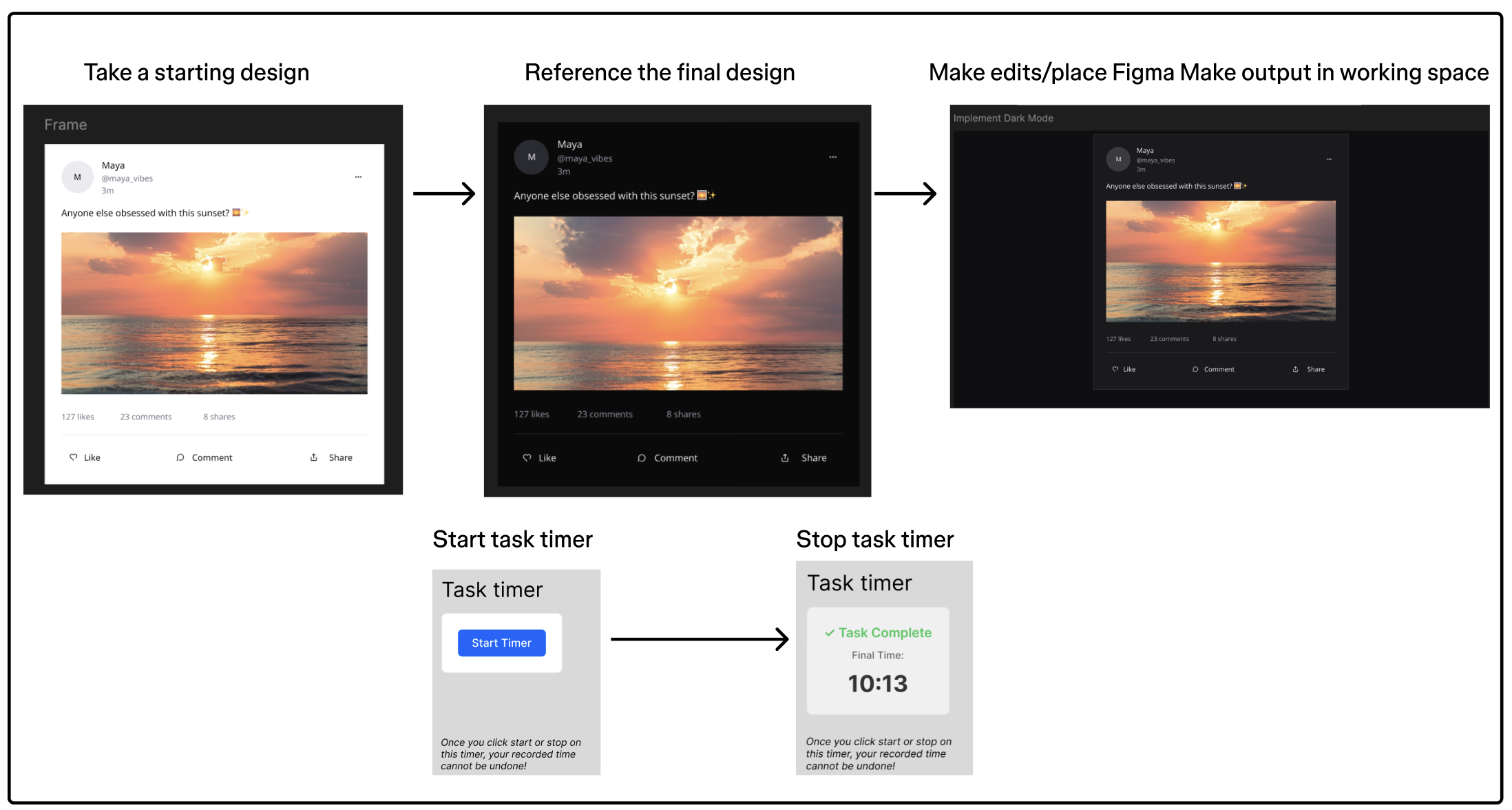}
  \caption{Experiment task flow for designing and time to completion measurement}
  \Description{A screenshot of reference designs and time-tracking widgets used in the study demonstrating participant workflow per unique study task.}
\end{figure}

Control participants were instructed to use the starting design frame and directly edit the design in the working space to match the reference image. Treatment participants were instead instructed to include the starting frame as an attachment alongside their initial prompt to Figma Make. They could send as many prompts to Figma Make as desired to edit the design.  Treatment participants were additionally allowed to return the design modified with Figma Make to the experiment file at any point to conduct manual edits. All participants were asked to place their updated designs in the task workspace when they believed they had successfully matched the reference image of the desired changes. This allowed moderators to compare the submitted designs with the reference images using task-specific completion criteria supplied by the study team.

Participants in both conditions attempted the same tasks and were evaluated against the same completion criteria. The sole difference was that treatment participants had access to Figma Make and no other Figma AI features. This was done in order to isolate Figma Make’s effect on task completion time. Participants in the treatment group were given five minutes to review a primer on Figma Make sourced from Figma Help Center articles. Participants were permitted to use external resources to complete each study task, including official Figma documentation. We conducted internal pilot trials of the study with product designers, product managers, designer advocates, and researchers, and revised the experiment multiple times based on their feedback.

\subsubsection{Moderation \& Data Collection}
All MeasuringU moderators were trained on how to administer the experiment by the internal study team, including following a standardized script and moderation documentation. Moderators walked participants through the instructions of each design task, reviewed each task’s final design to verify it matched the reference photo, and assisted participants with any troubleshooting errors when working through the scenarios. 

Time to completion for each task was recorded through a time-tracking Figma widget built for the study. Moderators started the timer at the participant's instruction and stopped the timer after verifying, according to the moderation guidelines, that the participant’s design matched the reference image. The widget additionally sent timestamped logs to an external database, which enabled the study team to cross-check the widget’s displayed completion times with separate telemetry.

After each task, participants rated how easy it was to complete the task on a seven-point scale from 1 ("very hard") to 7 ("very easy"). After completing all tasks, participants completed an exit survey collecting data on the perceived usability of Figma Make using the System Usability Scale (SUS) \cite{lewis2018}, self-reported design quality, and prior experience with AI tools. The questions asked in the exit survey can be found in Appendix B. 

\subsection{Data Analysis}
Data was collected through a combination of the experiment’s time-tracking widgets and supporting telemetry, task completion questions answered after each task, the eligibility screener, and the exit survey. We analyzed the collected data as follows. 

We employed generalized estimating equations (GEE) models to measure differences in completion rate and time to completion for each of the individual tasks as well as the cumulative time to complete all three scenarios. GEE models account for the correlations in repeated observations within participants rather than treating a given participant's performance on each task as independent measurements. By clustering per participant and adjusting standard errors to account for task outcome interdependence, we estimated both individual task performance as well as aggregate time to completion across the full trial. GEE models particularly excel at estimating the population-level average effects of time to task completion compared to alternative generalized linear model and mixed-effects models \cite{neuhaus1991}.

All GEE models included covariates for participant demographics, design work background, and AI experience to account for these characteristics when estimating the treatment effect. For selected categorical covariates, we combined categories with few participants to reduce the number of coefficients estimated. Completion times were log-transformed to address their right-skewed distribution. We assessed model assumptions regarding participant independence and residual normality. All models used cluster-robust sandwich standard errors \cite{liang1986} and are reported with 95\% confidence intervals. Average marginal effects (AMEs) and standard errors estimated by the models are reported in minutes and converted to relative percent differences. 

For categorical results regarding ease and usability scores as well as quality ratings, we employed Chi-squared testing as well as Mann–Whitney U tests as an alternative specification to test result robustness. Chi-squared results were additionally supplemented with adjusted standardized residuals to isolate the response categories contributing most to the observed differences along with the overall Chi-squared score \cite{greenwood1996}.

\section{Results}
\subsection{Descriptive Statistics}
A full review of descriptive statistics for the collected variables can be found in Appendix C. The sample was evenly split by gender. Most participants were aged 25–54 and had at least 10 years of professional experience. Participants primarily worked full time at organizations with 1–1,000 employees. The most common industries were information technology and banking and financial services. Most participants also worked with web-based applications and digital experiences. Product designers in the sample tended to have more professional experience, are older, and were more likely to be self-employed than product managers who participated in the study. 

Use of AI, Figma Design, and Figma AI features differs greatly between product designers and product managers. While the majority of both groups used AI tools daily, designers typically used AI for design work daily or a few times a week, whereas product managers generally used it once a week or less. The most popular AI tools across both job roles are ChatGPT, Google Gemini, and Claude. The strong majority of designers use Figma multiple times a day and have used Figma AI before, while product managers varied in how frequently they used Figma and most had not used Figma AI before. 

\begin{table}[t] \caption{Completion rates, time to completion, task ease scores, and System Usability Scale (SUS) scores. Values reported as means \& standard deviation.} \label{tab:performance} \centering \small \begin{tabular}{lccc} \toprule & Full Sample & Designers & PMs \\ \midrule \multicolumn{4}{l}{\textbf{Completion Rate}} \\ Task 1 (Dark Mode) & 93\% & 98\% & 88\% \\ Task 2 (Settings) & 91\% & 96\% & 86\% \\ Task 3 (Comment Modal) & 73\% & 90\% & 56\% \\ \addlinespace \multicolumn{4}{l}{\textbf{Time to Completion (minutes)}} \\ Task 1 (Dark Mode) & 11:23 (8:05) & 8:27 (5:11) & 14:25 (9:25) \\ Task 2 (Settings) & 9:58 (6:16) & 8:15 (5:54) & 11:46 (6:11) \\ Task 3 (Comment Modal) & 25:40 (9:13) & 24:29 (9:14) & 26:54 (9:08) \\ Tasks 1--3 (Cumulative) & 46:33 (15:14) & 41:13 (14:32)& 52:12 (13:59) \\ \addlinespace \multicolumn{4}{l}{\textbf{Task Ease Scores}} \\ Task 1 (Dark Mode) & 4.9 (1.7) & 5.7 (1.4) & 4.1 (1.5) \\ Task 2 (Settings) & 4.8 (1.7) & 5.3 (1.7) & 4.3 (1.7) \\ Task 3 (Comment Modal) & 3.4 (1.8) & 4.2 (1.7) & 2.7 (1.5) \\ \addlinespace \textbf{SUS Score} & 60.9 (2.2) & 70.1 (2.8) & 51.2 (2.7) \\ \bottomrule \end{tabular} \end{table}

\begin{table}[!b]
\centering
\caption{Full completion rate Treatment vs.\ Control GEE deltas across individual and average tasks. Values are percentage differences and standard errors.}
\label{tab:completion-rate-gee}
\small
\begin{tabular}{@{}lccc@{}}
\toprule
 & \textbf{Full Sample}
 & \makecell{\textbf{Product}\\\textbf{Designers}}
 & \makecell{\textbf{Product}\\\textbf{Managers}} \\
\midrule
\quad Task 1 (Dark Mode)
    & 0.8 (4.9)
    & $-$7.0 (4.9)
    & 8.6 (8.9) \\
\quad Task 2 (Settings)
    & 0.8 (5.1)
    & $-$10.9 (5.8)
    & 12.5 (8.9) \\
\quad Task 3 (Comment Modal)
    & 19.9\textsuperscript{**} (6.8)
    & 1.6 (7.9)
    & 38.1\textsuperscript{***} (11.4) \\
\quad Tasks 1--3 (average)
    & 7.2 (4.1)
    & $-$5.4 (4.5)
    & 19.7\textsuperscript{**} (7.4) \\
\bottomrule
\addlinespace[0.5em]
\multicolumn{4}{@{}l}{%
\footnotesize * $p < 0.05$, ** $p < 0.01$, *** $p < 0.001$.%
}
\end{tabular}
\end{table}

\subsection{RQ1 - Time Savings from Figma Make}
The average cumulative time to completion across all tasks for all participants was 46 minutes.  Completion rates and completion times varied substantially across tasks. Task 2 was the fastest task at an average of 10 minutes for all participants, while Task 3 where participants created a comment flyout interaction took the longest at an average of 26 minutes. Completion rates exceeded 90\% for Tasks 1 and 2, but fell to 73\% for Task 3. Treatment group participants submitted on average 2.9 Figma Make prompts for Task 1, 2.6 prompts for Task 2, and 5.2 prompts for Task 3. The complete distributions of individual task times for all participants can be found in Figure 2, and Table 1 summarizes completion rates and mean completion times.

\begin{figure}[t!]
  \centering
  \includegraphics[width=\linewidth]{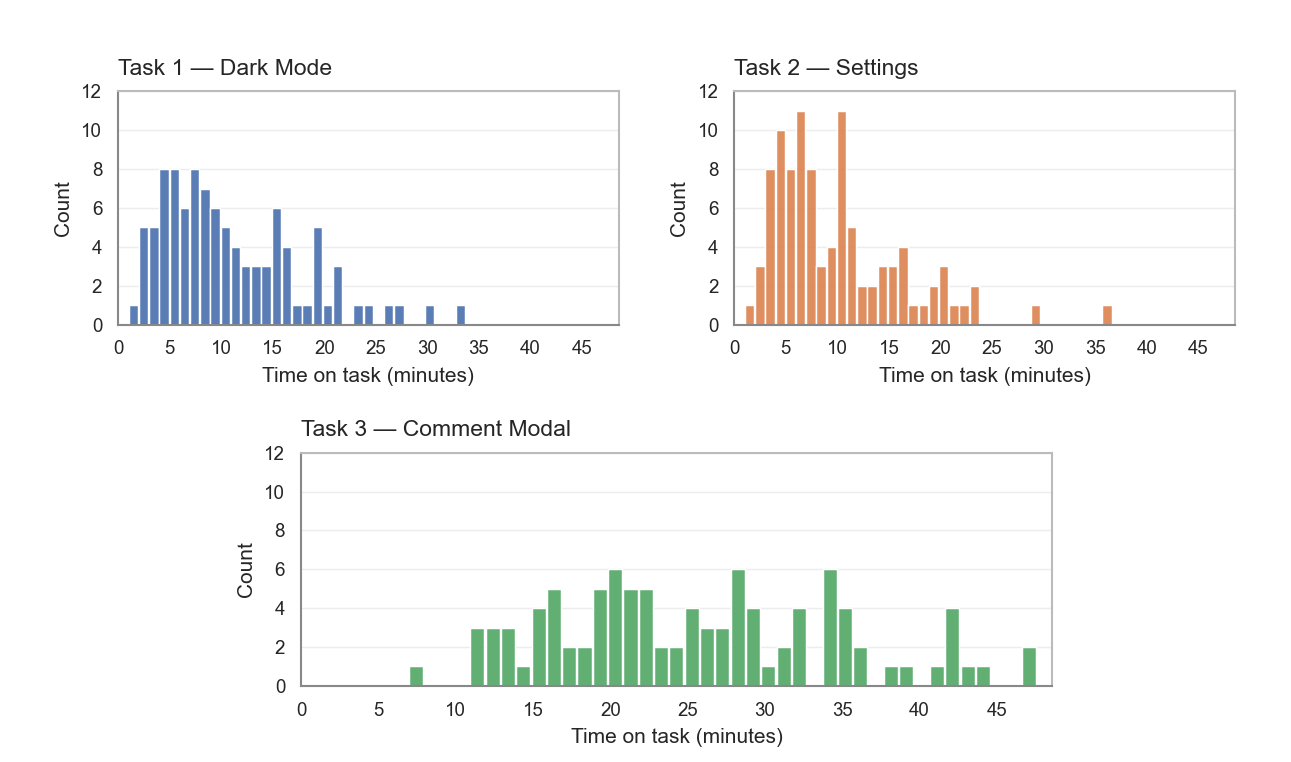}
  \caption{Distributions of study participant's time on each task }
  \Description{Three histogram charts of the time participants took on each of the three tasks featured within the experiment.}
\end{figure}

We first examined whether access to Figma Make affected participants’ likelihood of completing the tasks. In the analysis of completion probability, we did not find statistically significant completion differences for Tasks 1 or 2, or for the average across all three tasks (Table 2). We did however find that the estimated probability of completing Task 3 was 19.9 percentage points higher (SE=6.8\%, p=0.004) in the Figma Make condition than in the control condition 

Table 3 outlines treatment vs control deltas in minutes for participants who successfully completed each task. Cumulatively for the entire sample, having access to Figma Make versus no access to Make saves participants approximately 15 minutes of time, equivalent to a 27\% reduction in completion time (AME=15:08, SE=3.18, p<0.0001). Comparing the average of time saved across all tasks equally given the skewed role of Task 3 on cumulative time resulted in a 20\% reduction in time to completion (AME=10:53, SE=3:25, p=0.0014). Within the individual tasks, Tasks 2 and 3 both reported time savings improvements at 27\% (AME=2:52, SE=0:56, p=0.002) and 32\% (AME=11:21, SE=2:35, p<0.0001) relative time to task completion reductions respectively. Task 1 did not report a statistically significant effect in task time, but it is directionally reported at a 4.5\% relative time reduction.

We also examined associations between participant characteristics, task completion, and completion time. We have three key terms in our models - the independent variable of cumulative time to completion across the tasks, the primary dependent variable of treatment vs control group assignment, and the secondary dependent variable comparing product managers to product designers. All other controlled variable coefficients, standard errors, and significance test results can be referenced in Appendix D. 

Most demographic and AI experience variables report no statistical significance across both the completion rates and time to completion models. The primary exception to this trend is regarding users who never use AI tools in their work, which is only 2\% of the sample. Compared with participants who used AI tools for work a few times a week, these participants had an estimated completion rate 30.3 percentage points lower and an estimated cumulative completion time 35 minutes longer (Appendix D). Other lower signal coefficient results are that self employed and members of 10,000+ sized organizations had lower completion rates, that participants in communications, marketing, and media industries had higher completion rates, and that participants in information technology completed study tasks faster. 

\begin{table}[!b]
\centering
\caption{Time-to-completion Treatment vs.\ Control GEE minute deltas across individual and cumulative tasks. Values are population averages and standard errors.}
\label{tab:time-completion-gee}
\small
\begin{tabular}{@{}lccc@{}}
\toprule
 & \textbf{Full Sample}
 & \makecell{\textbf{Product}\\\textbf{Designers}}
 & \makecell{\textbf{Product}\\\textbf{Managers}} \\
\midrule
\quad Task 1 (Dark Mode)
    & 0:56 (1:01)
    & $-$1:21 (1:06)
    & 3:13 (1:43) \\
\quad Task 2 (Settings)
    & 2:52\textsuperscript{**} (0:56)
    & 1:10 (1:00)
    & 4:34\textsuperscript{**} (1:35) \\
\quad Task 3 (Comment Modal)
    & 11:21\textsuperscript{***} (2:35)
    & 7:41\textsuperscript{**} (2:47)
    & 15:00\textsuperscript{***} (4:33) \\
\quad Tasks 1--3 (cumulative)
    & 15:08\textsuperscript{***} (3:11)
    & 7:30\textsuperscript{*} (3:49)
    & 22:46\textsuperscript{***} (5:17) \\
\bottomrule
\addlinespace[0.5em]
\multicolumn{4}{@{}l}{%
\footnotesize * $p < 0.05$, ** $p < 0.01$, *** $p < 0.001$.%
}
\end{tabular}
\end{table}

\begin{table}[t]
\centering
\caption{GEE key contrasts on cumulative completion time. Values are population averages and standard errors in minutes.}
\label{tab:gee-key-coefficients}
\small
\setlength{\tabcolsep}{4pt}
\begin{tabular}{@{}lcc@{}}
\toprule
 & \textbf{Baseline} & \textbf{Interaction} \\
\midrule
Treatment
  & $-$10:30\textsuperscript{**} (3:18)
  & $-$3:03 (3:45) \\
\addlinespace[0.35em]
\multicolumn{3}{@{}l}{\textit{Job Role (Reference: Product Designer)}} \\
\quad Product Manager
  & 17:17\textsuperscript{***} (3:37)
  & 26:30\textsuperscript{***} (5:49) \\
\addlinespace[0.35em]
\multicolumn{3}{@{}l}{\textit{Treatment $\times$ Job Role}} \\
\quad Treatment $\times$ Product Designer
  & --
  & $-$3:03 (3:45) \\
\quad Treatment $\times$ Product Manager
  & --
  & $-$21:58\textsuperscript{***} (5:47) \\
\bottomrule
\addlinespace[0.5em]
\multicolumn{3}{@{}l}{%
\footnotesize * $p < 0.05$, ** $p < 0.01$, *** $p < 0.001$.%
}
\end{tabular}
\end{table}

Table 4 presents estimates from models with and without a treatment-by-role interaction. Within the baseline non-interacted GEE model, we observe a 20\% time reduction impact (AME=10:30, SE=3:18, p=0.0014) when removing the influence of job role into its own term at a 10 minutes and 30 seconds reduction. This estimate was similar in magnitude to the averaged time to completion in the fully saturated GEE model outlined in Table 3. Said replication across both approaches corroborates an approximately 20 percentage magnitude as the effect size for Figma Make's impact on time to task completion for the collective population of product designers and product manager within this study. 

\subsection{RQ2 - Designers versus Non-Designers Time Savings}
Completion rates across tasks sizably vary between product designers compared to product managers, with completion per task consistently above 90\% for designers, but between 86-88\% for product managers on Tasks 1 and 2 and dropping to 56\% on Task 3. The average cumulative time to completion across all tasks for product designers is 41 minutes, while product managers reported an average cumulative time of 52 minutes. The only significant differences in completion rates across the tasks comes from product managers and is primarily driven by the results of Task 3. Being in the treatment group raised PM completion of Task 3 by 38\% (SE=11.4\%, p=0.0008), producing a 20\% average completion rate lift for this group (SE=7.4\%, p=0.007). 

The estimated differences in completion time per task and cumulatively were largest among product managers. Being in the treatment group saved cumulatively 22 minutes and 46 seconds for PMs within the study, equating to a 35\% reduction in full experiment time to completion (SE=5:17, p<0.0001). Said gains were predominantly driven by Task 3 (AME=15:00, SE=4:33, p=0.001) and also supported by Task 2 (AME=4:34, SE=1:35, p=0.004). For product designers, improvements were only observed for Task 3 with a 7 minute and 41 second time reduction equivalent to a 26\% improvement (SE=2:47, p=0.006). These gains drove the overall time reduction across the tested workflow for product designers by a marginally significant 17\% (AME=7:30, SE=3:49, p=0.049).

In the model without interaction terms (Table 4), we observed a statistically significant 17 minute and 17 second coefficient for product managers compared to product designers as the reference group (SE=3:37, p<0.0001). This translates to relatively 39\% more minutes in cumulative task completion in the study solely from being a product manager compared to product designers. 

Interacting treatment assignment with job roles enables further exploration for how Make access contributed to time to completion at different magnitudes for product designers and product managers. The impact for product managers on being assigned to the treatment group in task time to completion is statistically significant at a 21 minute and 58 second reduction, equating to a 33\% relative time reduction when compared to product managers in the control group (SE=5:47, p=0.0001). This implies that having Make access almost makes up for product manager’s comparative time to completion disadvantage in tested design work compared to product managers without Make access. 

Reductions in time to completion is also observed for treatment group assigned product designers compared to control group product managers, but this term is not statistically significant. This observationally implies an 8\% reduction in time to completion for treatment group product designers, but the lack of statistical significance denotes that we cannot conclusively say that this result is not due to random chance alone. 

\subsection{RQ3 - Design Quality, Task Ease, and System Usability}
To address RQ3, we examined task-ease ratings collected after each task and exit-survey responses on perceived usability and design quality.  Ease scores were collected on a 1 to 7 Likert scale where 1 rated the task as “very difficult” and 7 rated that task as “very easy”.  Task ease scores closely matched the observed trends in time to completion.  Across both conditions, mean task-ease ratings were similar for Task 1 (M=4.91) and Task 2 (M=4.83), while Task 3 received lower ratings (M=3.44). Product managers reported lower mean ratings than product designers on all three tasks, with differences of 1.50, 0.98, and 1.36 points, respectively. 

The mean System Usability Scale (SUS) score was approximately 61 on a 0–100 scale. This score fell between the means associated with “OK” (50.9) and “Good” (71.4) in Bangor et al.’s adjective-rating SUS study \cite{bangor2009}. As with task-ease ratings, mean SUS scores were higher among product designers (70.1) than product managers (51.2)

In the exit survey, participants were asked questions regarding their perceived quality of their design work on all tasks, as well as whether Figma Make successfully assisted treatment group participants. Participants primarily rated their designs as somewhat matching the instructions for all experiment tasks, and they believed that the quality of their designs was either about the same or somewhat worse than what they would create outside of the experiment. Among treatment participants, 78\% (39/50) somewhat or strongly agreed that Figma Make would increase their ability to contribute to the design process. When asked whether they trusted Figma Make to interpret their intentions correctly, 42\% (21/50) somewhat agreed, while 32\% (16/50) neither agreed nor disagreed.

\begin{table}[!t]
\centering
\caption{Ease \& System Usability Scale (SUS) scores hypothesis test results. Values are average score deltas and standard errors.}
\label{tab:ease-sus-hypothesis}
\begin{tabular}{@{}lccc@{}}
\toprule
 & \textbf{Full Sample} & \textbf{Product Designers} & \textbf{Product Managers} \\
\midrule
\multicolumn{4}{@{}l}{\textit{Task Ease}} \\
\quad Task 1 (Dark Mode)     & 0.18 (0.66)                       & -0.65 (0.78)                     & 0.99\textsuperscript{*} (0.84) \\
\quad Task 2 (Settings)      & 1.18\textsuperscript{***} (0.65)  & 0.57 (0.94)                      & 1.77\textsuperscript{***} (0.81) \\
\quad Task 3 (Comment Modal) & 0.92\textsuperscript{**} (0.69)   & 0.31 (0.98)                      & 1.50\textsuperscript{***} (0.79) \\
\quad Tasks 1 + 2 + 3        & 0.76\textsuperscript{**} (0.57)   & 0.07 (0.77)                      & 1.42\textsuperscript{***} (0.61) \\
\addlinespace[0.5em]
\multicolumn{4}{@{}l}{\textit{SUS Score}} \\
\quad SUS Score              & 10.0\textsuperscript{*} (8.40)    & 2.94 (11.4)                      & 16.6\textsuperscript{**} (10.0) \\
\bottomrule
\addlinespace[0.5em]
\multicolumn{4}{@{}l}{\footnotesize\textit{Note.} * $p < 0.05$, ** $p < 0.01$, *** $p < 0.001$.}
\end{tabular}
\end{table}

Table 5 reports between-condition differences in task-ease and System Usability Scale (SUS) scores. Task ease reported statistically significant improvements on two out of the three of the tasks, leading to a 0.8 point average improvement on the 1-7 task ease scale (SE=0.29, p=0.009). SUS scores improved by a significant 10 points for treatment group participants on the score’s 0-100 scale (SE=4.23, p=0.02). These gains are equivalent to a 16\% relative improvement in ease and 15\% improvement in system usability enabled by Make access on the experiment tasks across all participants. 

Gains in ease and system usability ratings for participants are driven predominantly by product managers. All three tested tasks demonstrated significant improvements in ease for PMs with Make access than without, leading to an average relative percent improvement of 37\% (delta=+1.30 rated score, SE=0.32, p=0.0002). System usability via the SUS score increased by a relative 32\% for product managers (delta=+14.8 rated score, SE=5.18, p=0.006), which places the expected score for treatment group PMs at approximately the SUS scale average score. Neither task ease nor system usability reported any statistically significant changes for product designers. 

\begin{table*}[!b]
\centering
\caption{Quality ratings chi-squared test results.}
\label{tab:quality-ratings-chi-squared}
\begin{tabular}{@{}p{0.34\textwidth}p{0.20\textwidth}p{0.18\textwidth}p{0.20\textwidth}@{}}
\toprule
 & \textbf{Full Sample} & \textbf{Product Designers} & \textbf{Product Managers} \\
\midrule

``How closely do you believe your designs in this study matched the instructions that were provided in each scenario?''
    & 8.1\textsuperscript{*} \newline
      ``Slightly matched'' for control, ``Somewhat matched'' for treatment
    & 1.2 \newline
      --
    & 7.7\textsuperscript{*} \newline
      ``Slightly matched'' for control \\

\addlinespace[1em]

``How would you rate the quality of your created designs during this study compared to what you could have made in Figma outside of this study?''
    & 15.5\textsuperscript{**} \newline
      ``Much better'' for treatment, ``Much worse'' for control
    & 4.0 \newline
      --
    & 16.0\textsuperscript{**} \newline
      ``Much better'' for treatment, ``Much worse'' for control \\

\bottomrule
\addlinespace[0.5em]
\multicolumn{4}{@{}l}{\footnotesize\textit{Note.} * $p < 0.05$, ** $p < 0.01$, *** $p < 0.001$.}
\end{tabular}
\end{table*}

Finally, we compared responses to the two design-related exit-survey questions between the treatment and control conditions using chi-squared tests (Table 6). For both questions of how closely participants believed their designs matched the scenario instructions ($\chi^2(3)$ = 8.05, p=0.045, V=0.28), as well as how they would rate the quality of their experiment designs versus what they could have made outside of the study ($\chi^2(4)$ = 15.46, p=0.004, V=0.39), there is a statistically significant difference for treatment group participants compared to control. For the question of whether their design matched the scenario instructions, the control group’s greater likelihood of “slightly matched” (22\% vs 4\%) compared to the treatment group’s greater tendency towards “somewhat matched” (66\% vs 46\%) drives the chi-squared results. For the question of the quality of their designs within the study compared to what participants could have made outside of the experiment, a higher proportion of “much better” answers for treatment participants (20\% vs 2\%) and a higher proportion of “much worse” for control participants (14\% vs 2\%) is what primarily contributes to the chi-square term.

These findings are once again statistically significant for product managers, but they are not significant for product designers when considering job role breakouts. These results remain the same with post hoc correction tests. When considering an ordinal Mann-Whitney U test specification, we find the question of whether participants' designs matches the scenario instructions to no longer be significant in aggregate, but remaining significant for product managers specifically. In contrast, the result trends remain the same with ordinal testing regarding the quality of designs comparing within versus outside the experiment.

\section{Discussion}
\subsection{Synthesis of Results}
Estimated completion times for successfully completed tasks were approximately 20\% lower in the Figma Make condition than in the control condition when averaging across the three tasks. The estimates account for differences in demographic characteristics, professional experience, and AI tool familiarity. Participants in the Figma Make condition also reported higher task ease and perceived usability scores. These findings suggest that access to Figma Make may support increased productivity and ease of use in structured, reference-based design tasks. Whether these benefits extend to broader design work remains uncertain.

For product managers, access to Figma Make was associated with shorter completion times and higher task ease and perceived usability scores. Product managers in the treatment condition also rated their designs more favorably relative to what they thought they could produce outside the study. These findings suggest productivity benefits beyond time savings, encompassing perceived ease of use and self-assessed design quality. Moreover, prompt-to-design tools may help product managers translate their design intentions and ideas into interface changes. The gains observed among product managers, who generally used Figma less frequently in our sample, also are consistent with prior findings that AI assistance can benefit users with less experience or lower baseline performance \cite{noy2023, brynjolfsson2025, dellacqua2026}.

Comparatively, the results for product designers were less definitive. We did not find a statistically significant reduction in cumulative completion time among designers with access to Figma Make. However, designers in the treatment condition who completed Task 3, which required the creation of a comment flyout and was therefore the most complex and time intensive task, did so more quickly. This result supports the notion that prompt-to-design tools may offer benefits and improvements for more involved, time-intensive tasks within established design workflows, even when there isn't an overall time-savings. 

The broader implication of these findings is therefore that prompt-to-design tools may reduce the time and effort required for specific UI design tasks, supporting designers’ established workflows while lowering barriers to participation for non-designers.

\subsection{Limitations}
There are five central limitations to this study. First, our time-to-completion analysis was limited to successfully completed tasks. Because Figma Make also affected the likelihood that a participant would complete a given task, particularly for product managers on Task 3, the subset of participants who completed each task may differ between the treatment and control conditions. Therefore, the reported time reduction should be interpreted alongside the completion results.

Second, the experiment used three standardized tasks that participants attempted in a fixed order, each with a specified target design. While this structure allowed consistent measurement across each condition, it does not capture the ambiguity, need to encompass stakeholder feedback, and changing requirements that are common in professional design projects. Therefore, the observed differences in completion rates and completion times with Figma Make may not extend to different types of design tasks or tasks conducted in a different order.

Third, task completion was verified by moderators by comparing participants’ submitted designs with the reference images. Although moderators followed the same guidelines, human variability in moderation may still impact the final results. 

Fourth, participants encountered stochastic rendering failures and occasionally needed additional prompts before Figma Make produced an editable design. These failures resulted in increased completion times among treatment participants. Because this behavior existed within the version of Figma Make available during the study, it was included in all measurements. As a result, our reported differences in time to completion between treated and non-treated participants may underestimate the true effect of prompt-to-design tools.

Finally, the experiment population consisted of 100 participants who viewed or edited Figma files at least monthly and had at least two years of experience as either a product designer or a product manager. These findings therefore do not necessarily generalize to newer Figma users, other design roles, or users of other prompt-to-design systems. Furthermore, product managers were the only non-design role included; thus, it is unclear if these results would extend to a broader non-designer population.

\subsection{Future Directions}
While this study provides an initial causal time savings measurement provided by AI design tools within an RCT framework, there are several different areas that researchers can explore in future work. Future experiments should consider including larger and more varied sets of design tasks as well as evaluating open-ended assignments without a predetermined output sequence. Longitudinal field studies could examine whether initial gains persist as users develop and refine prompting strategies and whether time saved during initial production is offset by stakeholder review or later iterations.

Future studies should also incorporate expert assessments of design quality with predefined rubrics. These rubrics should include dimensions such as visual quality, accessibility, and design-system or brand alignment. This would provide evidence on whether creators’ assessments of design quality are supported by independent expert evaluations and whether increased productivity comes at the expense of design quality. Researchers could also extend this work to other prompt-to-design tools, product versions, and professional populations. Testing different tools and product versions would help establish the boundaries of the observed effects, while testing different professional populations would provide insight into how those effects vary across design-adjacent roles.

\section{Acknowledgments}
\begin{acks}
We would like to extend our gratitude to Caitlin Wang, John Doherty, Minami Rojas, Clancy Slack, Julia Kirkpatrick, Shane Johnston, Prasant Loukendi, Mallory Dean, Lauren Byrne, Wayne Lin, Taryn Cowart, Jackie Chui, Hauke Sandhaus, Tammy Tassembaum, Jiayan Yu, Rodrigo Davies, Anna Astrom, Alia Fite, Madeline Stafford, Emma Webster, Alex Praeger, and the team at MeasuringU for helping to make this study possible. 
\end{acks}

\bibliographystyle{ACM-Reference-Format}
\bibliography{references}

\clearpage
\appendix
\section{Scenario Designs}
\subsection{Task 1 - "Create an alternative version of the social media post in dark mode."}
\begin{figure}[H]
  \centering
  \includegraphics[width=\linewidth]{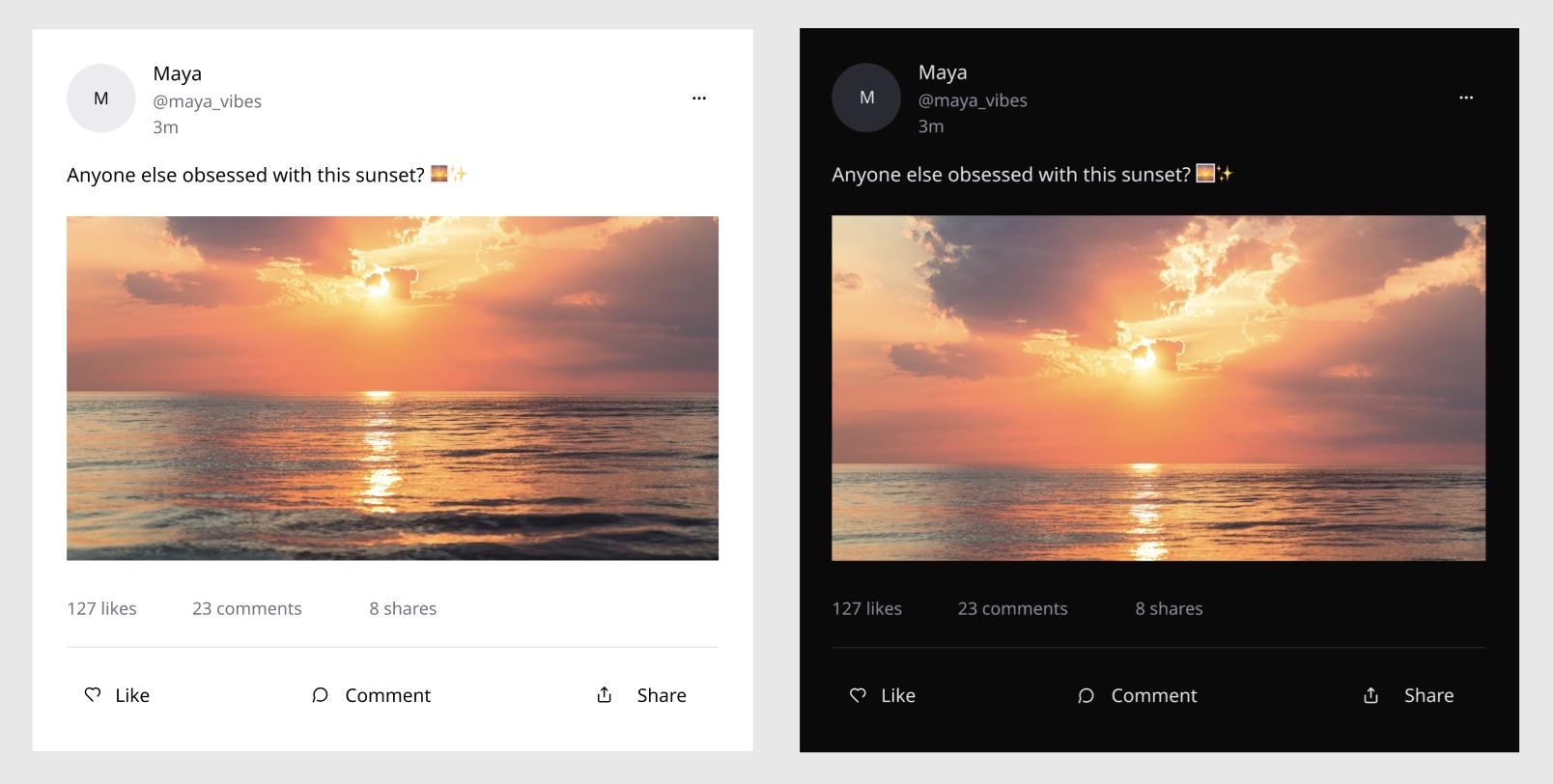}
  \Description{A screenshot of reference designs for the start of task 1 and the final edited version within the study.}
\end{figure}

\subsection{Task 2 - "Add a “Help \& Support” option to the settings drop down menu."}
\begin{figure}[H]
  \centering
  \includegraphics[width=\linewidth]{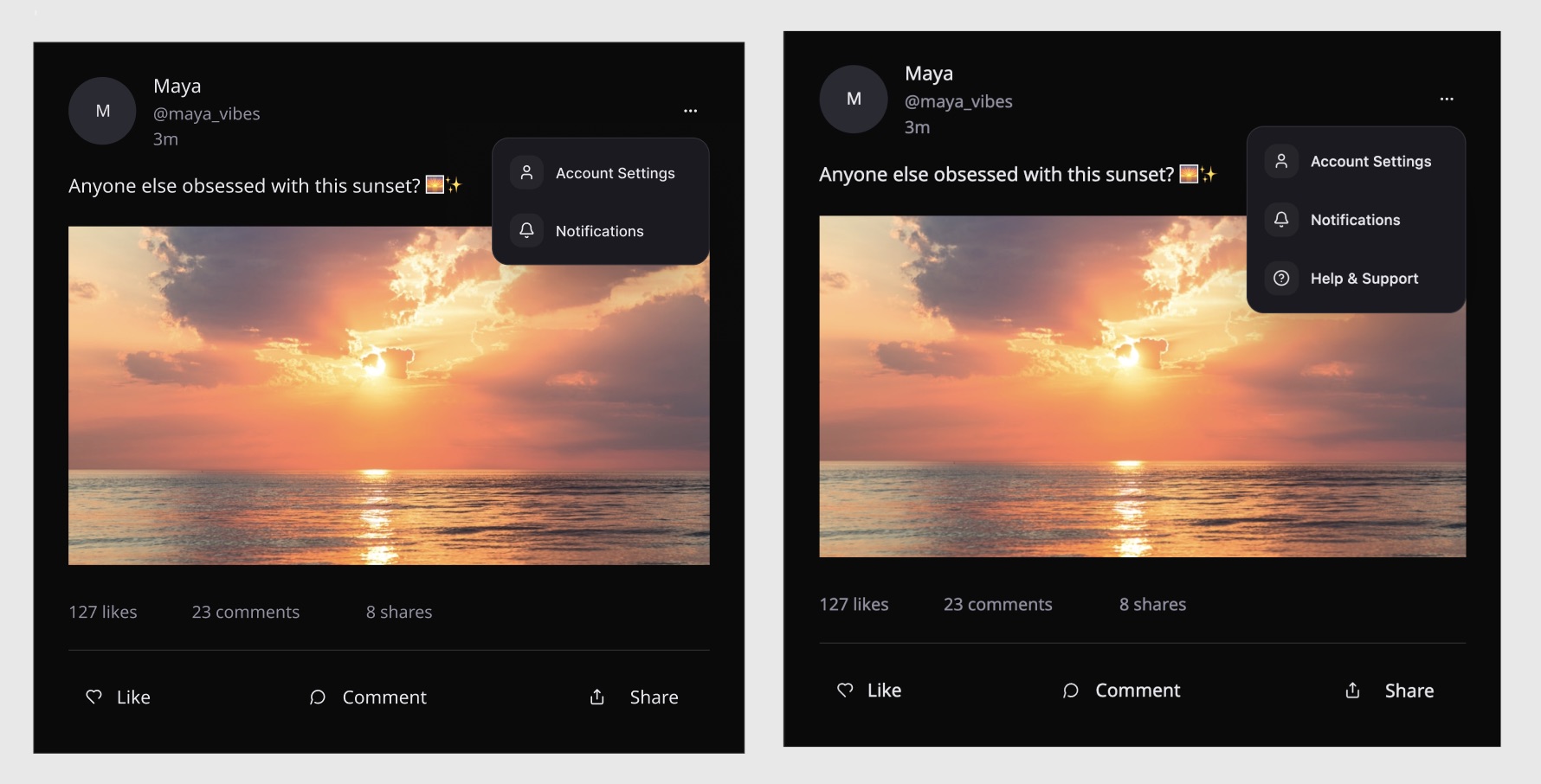}
  \Description{A screenshot of reference designs for the start of task 2 and the final edited version within the study.}
\end{figure}

\subsection{Task 3 - "Create a comment flyout generated from the bottom of the homepage after clicking 'Comment'”.}
\begin{figure}[H]
  \centering
  \includegraphics[width=\linewidth]{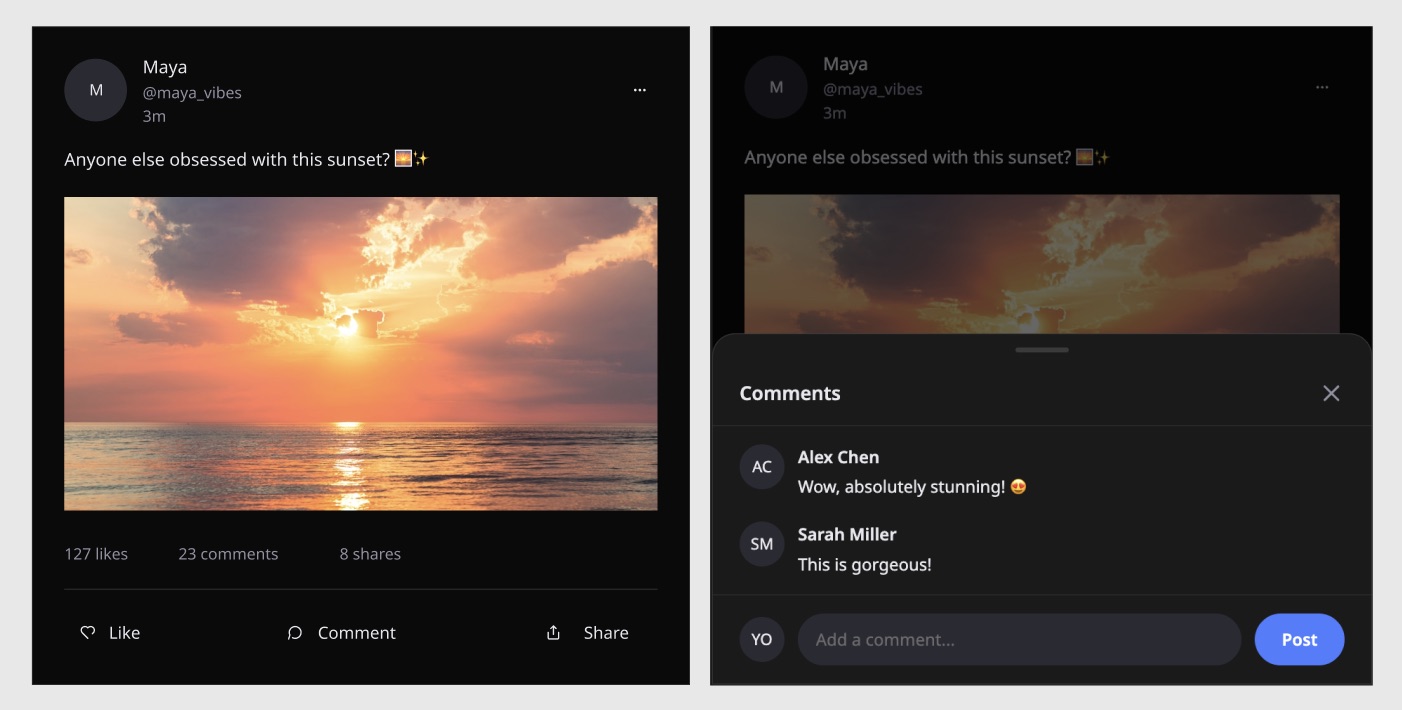}
  \Description{A screenshot of reference designs for the start of task 3 and the final edited version within the study.}
\end{figure}

\section{Exit survey questions}

\subsection{System Usability Questions}

\begin{enumerate}
  \item[\textbf{Q1.}] I think that I would like to use Figma frequently.
  \item[\textbf{Q2.}] I found Figma unnecessarily complex.
  \item[\textbf{Q3.}] I thought Figma was easy to use.
  \item[\textbf{Q4.}] I think that I would need the support of a technical person to be able to use Figma.
  \item[\textbf{Q5.}] I found the various functions in Figma were well integrated.
  \item[\textbf{Q6.}] I thought there was too much inconsistency in Figma.
  \item[\textbf{Q7.}] I would imagine that most people would learn to use Figma very quickly.
  \item[\textbf{Q8.}] I found Figma very cumbersome to use.
  \item[\textbf{Q9.}] I felt very confident using Figma.
  \item[\textbf{Q10.}] I needed to learn a lot of things before I could get going with Figma.
\end{enumerate}

\subsection{Design Quality Questions}

\begin{enumerate}
  \item[\textbf{Q11.}] How closely do you believe your designs in this study matched the instructions provided in each scenario?
  \item[\textbf{Q12.}] How would you rate the quality of your created designs during this study compared to what you could have made in Figma outside of this study?
  \item[\textbf{Q13.}] Have you ever used Figma Prototyping before this study?
\end{enumerate}

\subsection{Treatment-Only Questions}

\begin{enumerate}
  \item[\textbf{Q14.}] I trusted the AI tools available in this study to interpret my intentions correctly.
  \item[\textbf{Q15.}] I believe Figma AI would increase my ability to contribute to the design process.
\end{enumerate}

\subsection{AI Use Questions}

\begin{enumerate}
  \item[\textbf{Q16.}] How often do you use AI tools for work?
  \item[\textbf{Q17.}] How often do you use AI tools for design?
\end{enumerate}

\section{Participant Attribute Descriptive Statistics}
\begin{longtable}{lccc}
\caption{Participant demographics and background characteristics distributions.}
\label{tab:participant-demographics}\\
\toprule
Variable & Full Sample & Designers & PMs \\
\midrule
\endfirsthead
\caption[]{Participant demographics and background characteristics. Continued.}\\
\toprule
Variable & Full Sample & Designers & PMs \\
\midrule
\endhead
\midrule
\multicolumn{4}{r}{\textit{Continued on next page}}\\
\endfoot
\bottomrule
\endlastfoot
\multicolumn{4}{l}{\textit{Age}} \\
18--24 & 2\% & 4\% & 0\% \\
25--34 & 29\% & 24\% & 35\% \\
35--44 & 32\% & 35\% & 29\% \\
45--54 & 29\% & 35\% & 22\% \\
55--64 & 6\% & 0\% & 12\% \\
65+ & 2\% & 2\% & 2\% \\
\addlinespace
\multicolumn{4}{l}{\textit{Gender}} \\
Male & 50\% & 55\% & 45\% \\
Female & 50\% & 45\% & 55\% \\
Other & -- & -- & -- \\
\addlinespace
\multicolumn{4}{l}{\textit{Education}} \\
Undergraduate degree & 64\% & 73\% & 55\% \\
Graduate degree & 36\% & 27\% & 45\% \\
\addlinespace
\multicolumn{4}{l}{\textit{Employment Status}} \\
Employed full time & 78\% & 69\% & 88\% \\
Freelance / contractor & 16\% & 20\% & 12\% \\
Self-employed & 6\% & 11\% & -- \\
\addlinespace
\multicolumn{4}{l}{\textit{Years of Experience}} \\
2--5 years & 11\% & 8\% & 14\% \\
5--10 years & 29\% & 21\% & 37\% \\
10+ years & 60\% & 71\% & 49\% \\
\addlinespace
\multicolumn{4}{l}{\textit{Industry Sector}} \\
Information technology & 46\% & 35\% & 57\% \\
Banking or financial services & 20\% & 18\% & 22\% \\
Media, entertainment, arts & 9\% & 14\% & 4\% \\
Communications or marketing & 8\% & 16\% & -- \\
Retail & 5\% & 4\% & 6\% \\
Other & 12\% & 14\% & 10\% \\
\addlinespace
\multicolumn{4}{l}{\textit{Size of Organization}} \\
1--1,000 employees & 55\% & 62\% & 47\% \\
1,000--10,000 employees & 21\% & 14\% & 25\% \\
10,000+ employees & 24\% & 24\% & 28\% \\
\addlinespace
\multicolumn{4}{l}{\textit{Frequency of Figma Design Use}} \\
Multiple times a day & 55\% & 73\% & 37\% \\
Once a day & 14\% & 8\% & 20\% \\
Once a week & 18\% & 14\% & 22\% \\
Once a month or less & 13\% & 6\% & 20\% \\
\addlinespace
\multicolumn{4}{l}{\textit{Have Ever Used Figma AI}} \\
Yes & 49\% & 61\% & 37\% \\
No & 51\% & 39\% & 63\% \\
\addlinespace
\multicolumn{4}{l}{\textit{Frequency of AI Tool Use for Work}} \\
Daily & 78\% & 69\% & 88\% \\
A few times a week & 15\% & 20\% & 10\% \\
Once a week & 5\% & 8\% & 2\% \\
Once a month & 0\% & -- & -- \\
Never & 2\% & 4\% & -- \\
\addlinespace
\multicolumn{4}{l}{\textit{Frequency of AI Tool Use for Design}} \\
Daily & 15\% & 26\% & 4\% \\
A few times a week & 35\% & 45\% & 25\% \\
Once a week & 20\% & 12\% & 29\% \\
Once a month & 13\% & 10\% & 16\% \\
Never & 17\% & 8\% & 27\% \\
\addlinespace
\multicolumn{4}{l}{\textit{Previously Used AI Tools}} \\
Figma Make & 34\% & 43\% & 25\% \\
ChatGPT & 94\% & 92\% & 96\% \\
Lovable & 34\% & 26\% & 43\% \\
Google Gemini & 79\% & 77\% & 82\% \\
Bolt & 11\% & 12\% & 10\% \\
GitHub Copilot & 26\% & 18\% & 35\% \\
Claude & 72\% & 65\% & 80\% \\
Replit & 14\% & 10\% & 18\% \\
Cursor & 24\% & 20\% & 29\% \\
\end{longtable}

\section{Regression model covariate terms}

\begin{longtable}{@{}lcc@{}}
\caption{GEE covariate coefficients for completion and cumulative time}
\label{tab:gee-covariates} \\
\toprule
 & \textbf{Completion (pp)} & \textbf{Cumulative time} \\
\midrule
\endfirsthead
\multicolumn{3}{c}{\tablename\ \thetable{} -- continued from previous page} \\
\toprule
 & \textbf{Completion (pp)} & \textbf{Cumulative time} \\
\midrule
\endhead
\midrule
\multicolumn{3}{r}{\footnotesize Continued on next page} \\
\endfoot
\bottomrule
\addlinespace[0.5em]
\multicolumn{3}{@{}l}{%
\footnotesize * $p < 0.05$, ** $p < 0.01$, *** $p < 0.001$.%
} \\
\endlastfoot
\multicolumn{3}{@{}l}{\textit{Age (Reference: 18--34)}} \\
\quad 35--44
  & $-$1.4 (4.6)
  & $-$0:57 (4:22) \\
\quad 45--54
  & $-$3.5 (6.2)
  & 6:00 (6:02) \\
\quad 55+
  & $-$12.3 (11.1)
  & 13:27 (10:01) \\
\addlinespace[0.35em]
\multicolumn{3}{@{}l}{\textit{Gender (Reference: Male)}} \\
\quad Female
  & 3.5 (4.3)
  & 0:08 (3:01) \\
\addlinespace[0.35em]
\multicolumn{3}{@{}l}{\textit{Education (Reference: Undergraduate)}} \\
\quad Graduate degree
  & $-$6.1 (3.9)
  & $-$1:36 (3:29) \\
\addlinespace[0.35em]
\multicolumn{3}{@{}l}{\textit{Employment Status (Reference: Freelance/Contractor)}} \\
\quad Full Time
  & 3.1 (4.7)
  & $-$2:31 (4:33) \\
\quad Self Employed
  & $-$20.8\textsuperscript{**} (7.0)
  & 6:24 (8:31) \\
\addlinespace[0.35em]
\multicolumn{3}{@{}l}{\textit{Years of Experience (Reference: 2--5 years)}} \\
\quad 5--10 years
  & $-$13.8\textsuperscript{*} (5.7)
  & 3:42 (5:06) \\
\quad 10+ years
  & $-$9.5 (6.9)
  & 8:22 (6:19) \\
\addlinespace[0.35em]
\multicolumn{3}{@{}l}{\textit{Industry (Reference: Banking \& financial services)}} \\
\quad Information technology
  & 11.3 (5.7)
  & $-$8:36\textsuperscript{*} (4:20) \\
\quad Communications \& Marketing
  & 13.8\textsuperscript{*} (6.4)
  & $-$6:30 (4:46) \\
\quad Media \& Entertainment
  & 13.5\textsuperscript{*} (6.6)
  & $-$0:26 (5:41) \\
\quad Retail
  & 15.3 (7.9)
  & $-$12:39\textsuperscript{*} (5:35) \\
\quad Other
  & $-$1.6 (8.4)
  & $-$1:09 (4:37) \\
\addlinespace[0.35em]
\multicolumn{3}{@{}l}{\textit{Organization Size (Reference: 1--1,000 employees)}} \\
\quad 1,000--10,000 employees
  & $-$9.3 (5.0)
  & 2:13 (3:26) \\
\quad 10,000+ employees
  & $-$12.6\textsuperscript{**} (4.8)
  & 5:21 (4:21) \\
\addlinespace[0.35em]
\multicolumn{3}{@{}l}{\textit{Figma Design Use Frequency (Reference: Few times a year or less)}} \\
\quad Multiple times a day
  & 4.9 (6.7)
  & 3:18 (5:09) \\
\quad Once a day
  & 3.9 (8.3)
  & 1:15 (5:30) \\
\quad Once a week
  & $-$4.8 (8.2)
  & 1:36 (5:37) \\
\addlinespace[0.35em]
\multicolumn{3}{@{}l}{\textit{Used Figma AI (Reference: Have not used)}} \\
\quad Have used
  & 5.7 (4.7)
  & $-$6:43\textsuperscript{*} (3:13) \\
\addlinespace[0.35em]
\multicolumn{3}{@{}l}{\textit{AI Tools For Work Frequency (Reference: Few times a week)}} \\
\quad Daily
  & 4.4 (3.9)
  & $-$3:49 (3:57) \\
\quad Once a week
  & $-$3.4 (12.4)
  & 0:06 (7:57) \\
\quad Never
  & $-$30.3\textsuperscript{**} (10.9)
  & 35:00\textsuperscript{**} (13:08) \\
\addlinespace[0.35em]
\multicolumn{3}{@{}l}{\textit{AI Tools for Design Frequency (Reference: Few times a week)}} \\
\quad Daily
  & $-$8.4 (7.1)
  & $-$4:14 (4:28) \\
\quad Once a week
  & 16.8\textsuperscript{*} (7.0)
  & $-$0:37 (4:18) \\
\quad Once a month
  & $-$1.9 (7.9)
  & 5:58 (7:45) \\
\quad Never
  & 0.3 (7.7)
  & $-$4:11 (4:27) \\
\end{longtable}

\end{document}